\documentclass[sigconf, nonacm]{acmart}
\usepackage{algorithm}
\usepackage{algpseudocode}
\usepackage{enumitem}

\begin{document}
\sloppy 

\thanks{This research is supported by 
NSF$\#$1610282
"DataStorm: A Data Enabled System for End-to-End Disaster Planning and Response",
NSF$\#$2125246 "PanCommunity: Leveraging Data and Models for Understanding and Improving Community Response in Pandemics", $\#$2200140 ”Predicting Emergence in Multidisciplinary Pandemic Tipping-points (PREEMPT)”, and EU$\#$955708 "EvoGamesPlus: Evolutionary Game Theory and Population
Dynamics"}

\title{DataStorm-EM: Exploration of Alternative Timelines within Continuous-Coupled Simulation Ensembles}

\author{Fahim Tasneema Azad
}
\affiliation{%
  \institution{Arizona State University}
}

\author{Javier Redondo Anton}
\affiliation{%
  \institution{University of Torino}
}

\author{
Shubhodeep Mitra,
Fateh Singh,
Hans Behrens,
Mao-Lin Li,
Bilgehan Arslan, K. Sel\c{c}uk Candan}
\affiliation{%
  \institution{Arizona State University}
}

\author{Maria Luisa Sapino}
\affiliation{%
  \institution{University of Torino}
}

\begin{abstract}
Many socio-economical critical domains (such as sustainability, public health, and disasters) are characterized by highly complex and dynamic systems, requiring data and model-driven simulations to support decision-making. Due to a large number of unknowns, decision-makers usually need to generate ensembles of stochastic scenarios, requiring hundreds or thousands of individual simulation instances, each with different parameter settings corresponding to distinct scenarios,
As the number of model parameters increases, the number of potential timelines one can simulate increases exponentially. Consequently, simulation ensembles are inherently sparse, even when they are extremely large. This necessitates a platform for (a) deciding which simulation instances to execute and (b) given a large simulation ensemble, enabling decision-makers to explore the resulting alternative timelines, by extracting and visualizing consistent, yet diverse timelines from continuous-coupled simulation ensembles.
In this article, we present {\em DataStorm-EM} platform for data- and model-driven simulation ensemble management, optimization, analysis, and exploration, describe underlying challenges and present our solutions.
\end{abstract}

\maketitle

\section{Introduction}

The ability to make informed forecasts is critical to operating effectively, efficiently, safely, and securely in many contexts. Many socio-economical critical human-centered domains (such as sustainability, public health, and disasters) are characterized by highly complex and dynamic systems, requiring data and model-driven simulations to support decision-making. 
Yet, simulation-driven decision-making is challenging due to various reasons. Models often involve 100s of inter-dependent parameters, spanning multiple layers, communities, and geospatial frames, affected by complex, heterogeneous processes operating at different resolutions. Secondly, due to the unpredictability of the underlying process (such as a hurricane or a pandemic)  and the unpredictability of the actions of various independent agencies, decision-makers need to generate ensembles of many thousands of simulations, each with different parameters corresponding to plausible scenarios. These lead two several major research directions:
\begin{itemize}[leftmargin=*]
\item {Which simulation instances to include in the ensemble?}
\item {How to efficiently execute continuous coupled simulations?}
\item {How to interpret simulation ensembles consisting of complex multi-variate time series data sets that are hard to visualize.  }
\end{itemize}
As detailed below, we addressed the first two questions in previous work. This demonstration focuses on the \underline{third} challenge.
\begin{figure}[t]
    \centering
    {\includegraphics[width=.4\textwidth]{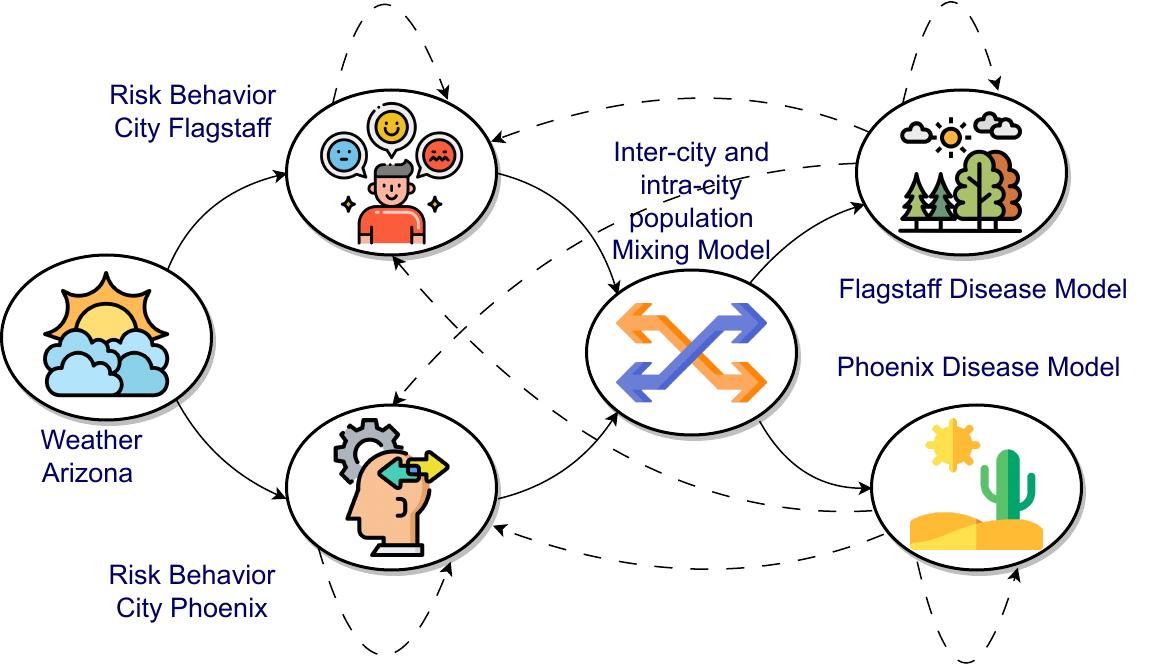}}
        \vspace*{-0.2in}
    \caption{A sample workflow that takes into account disease models, external impactors (such as weather), differences in local behaviors (such as {\em risk averseness}), and intra- and inter- city mixing patterns impacted by these behaviors}
    \label{fig:modeldiagram}
    \vspace*{-0.1in}
\end{figure}

\subsection{Preliminary Work: DataStorm-FE}
In~\cite{datastorm, behrens2018datastorm}, we introduced the {\em DataStorm Flow Engine} (DataStorm-FE) multi-model plug-and-play ensemble simulation platform for the creation, storage, and analysis of coupled, multi-model simulation ensembles. DataStorm-FE aims to help tackle the computational challenges underlying complex heterogeneous system modeling through advanced seamless integration of independently developed, reusable model and analysis components into a continuous-coupled simulation workflow (Figure~\ref{fig:modeldiagram}).
{\em DataStorm-FE} supports (a) model/data integration and alignment, (b) distributed/parallel workflow orchestration,  and (c) provides an adaptive orchestration layer (using Ansible to automate configuration management and Vagrant for provisioning and deployment) capable of creating, configuring, and destroying instances as needed to balance performance and budget requirements.
The platform coordinates data and decision flows among multiple simulation engines (each modeling a different aspect of a complex system) and enables end-to-end ensemble planning and optimization (including parameter-space sampling, output aggregation and alignment, and state and provenance data management) to improve the predictive accuracy of the overall end-to-end simulation process within a limited simulation budget. 
These are achieved through innovations in model/data integration and alignment~\cite{datastorm}, multi-model and multi-scale simulation ensemble creation
\cite{behrens2018datastorm, azad2022sirtem,icbk}, large-scale, high dimensional optimization~\cite{golf,wsc}, and multi-variate time series analysis~\cite{xm2a,selego}.

\subsection{Contributions of this Work: DataStorm-EM}
\label{ssec:EM}
In this work, we focus on the DataStorm's ensemble management, analysis, and exploration framework, {\em DataStorm-EM}., which complements the {\em DataStorm-FE} flow engine which creates continuous coupled simulation ensembles. 

As the number of model parameters of a simulation increases, the number of potential situations one can simulate increases exponentially -- consequently, simulation ensembles are inherently sparse, even when they are extremely large. 
Naturally, simulation ensembles generated by {\em DataStorm-FE} flow engine are only as useful as the conclusions they produce. As the raw multi-variate time series data generated by DataStorm-FE can be both large and complicated to parse, and users may be domain experts and not necessarily data scientists, it is critical that data exploration and analysis be accessible: 
\begin{itemize}[leftmargin =*]
    \item Consequently, {\em DataStorm-EM} provides intuitive and easy-to-use processes that allow users to explore complex alternative timelines of potential futures encapsulated by a simulation ensemble.
\end{itemize}
Our contribution contrasts to (and complements) more traditional complex event processing and stream processing systems: our goal in developing DataStorm-EM is not to efficiently process data streams (which is handled by DataStorm-FE), but rather to help (a) decision makers to decide which simulation instances to execute and (b) given a large simulation ensemble, consisting of multi-variate time series, to enable decision makers to explore the alternative (consistent, yet diverse) timelines captured by the ensemble.

\section{DataStorm-EM}
As we described above, {\em DataStorm-EM} complements the {\em DataStorm-FE} flow engine~\cite{behrens2018datastorm}, which implements and executes {\em coupled} and {\em continuous} simulation workflows, consisting of multiple models, each representing a different aspect of a complex system (Figure~\ref{fig:modeldiagram}).

\subsection{Background}

\subsubsection{DataStorm Workflows, Models, and Streams}
Inputs to a DataStorm flow are the data sources, with each step in the decision flow involving an analytic function, model, or decision criterion. Inputs to each step are data, decisions from previous steps, and user-provided decision parameters. The resulting data and decisions are outputs. 
A DataStorm workflow is defined as follows:
	
	\begin{definition}[DS-Flow]
		A DataStorm workflow is a node-labeled directed graph, $G(V,E,l)$, where the label, $l(v_h)$, of node $v_h \in V$ in the graph is a model, $\mathbb{M}_h$ -- i.e, $l(v_h) = \mathbb{M}_h$.  
	\end{definition}  
	
	\begin{definition}[Model]
		A model, $\mathbb{M}$, is a tuple $\langle F, \mathcal{P}, \mathcal{D}, \mathcal{O},\omega_I,\omega_O, \Delta\rangle$, where $F$() is a (simulation or analysis) function, $\mathcal{P}$ is a set of simulation parameter names (defining a simulation space), $\mathcal{D}$ is a set of input data variables, and $\mathcal{O}$ is a set of output variables, such that $\vec{o} \leftarrow F(\vec{p},\vec{d}),$
		where $\vec{p}$, $\vec{d}$, and $\vec{o}$ are instantiation vectors for $\mathcal{P}$, $\mathcal{D}$, $\mathcal{O}$, respectively.
	\end{definition}

\noindent	Input and output variables to $F$\textit{()} have temporal scopes:
The input variables capture input data to the model for a given input window size, $w_{in}$ (i.e., the temporal input scope is $[t,t+w_{in})$) for some time instant $t$; note that in discrete simulations, the input data will also have a discrete resolution, given by $res_{in}$.
The output variables, on the other hand, capture simulation/model output for a given window size, $w_{out}$. (i.e., the temporal output scope is $[t, t + w_{out})$) for some time $t$; in discrete simulations, the output data will also have a discrete resolution, given by $res_{out}$.
	In the definition of the model, $\omega_I = \langle w_{in}, res_{in}\rangle $ and $\omega_O= \langle w_{out}, res_{out}\rangle$ are the descriptors of the input and output scopes, e.g. the temporal lengths and resolutions for inputs and outputs.
	$\Delta$ is a the value of temporal shift.

Due to the temporal nature of the models,  DataStorm models  consume and produce data as streams:

               	\begin{definition}[Input Stream]
		Let $\mathbb{M}$ be a model. We refer to the sequence ${I}=(\vec{d}_{(0)}, \vec{d}_{(1)}, \vec{d}_{(2)}\ldots)$, where each $\vec{d}_{(j)}$ is defined in the input data space, $\mathcal{D}$, as an input stream.
	\end{definition}
	
	\begin{definition}[Output Stream]
		Let $\mathbb{M}$ be a model. Let ${I}$ be an input stream and $\vec{p}$ be a parameter instantiation. We refer to the sequence ${O}=(\vec{o}_{(0)}, \vec{o}_{(1)}, \vec{o}_{(2)}\ldots)$, where each $\vec{o}_{(j)}$ is defined in the output data space, $\mathcal{O}$, as the corresponding output stream.
	\end{definition}

	\subsubsection{DS-Actors and Local Ensembles}\label{sec:ds-flows}

In DataStorm, models are implemented within software modules called \textit{DS-Actors}. Ensemble creation involves coupled continuous execution, where 
each actor is invoked repeatedly consuming data produced upstream (in time or in workflow) actors and producing new data for downstream  (in time or in workflow) actors.
	\begin{definition}[Execution Step of a DS-Actor]\label{def:step}
		Let $G(V,E,l)$ be a DS-Flow, let $v \in V$ be a DS-Actor, and let $\mathbb{M} = \langle F, \mathcal{P}, \mathcal{D}, \mathcal{O}, \omega_I, \omega_O,\Delta\rangle$ be the corresponding model.
		Each execution step $s$ for DS-Actor $v$ has a corresponding input and output window, $in(s)$ and $out(s)$
		\begin{itemize}[leftmargin=*]
			\item if $s$ is a source node in $G$, then $in(s) = \bot$ and $out(s) = [s\times \Delta,  s\times \Delta +  w_{out})$;
			\item if $s$ is not a source node, then $in(s) = [s\times \Delta,  s\times \Delta + w_{in})$ and $out(s) = [s\times \Delta,  s\times \Delta + w_{out})$, where $\Delta$ is the value of the shift parameter and  $w_{in}$ and $w_{out}$ are the input and output window sizes defined by $\omega_I$ and $\omega_O$, respectively. 
		\end{itemize}    
	\end{definition}

\noindent	
DataStorm
	provides {\em stateless} and {\em stateful}
 models.

An important function of the DS-Actor is to sample vectors, $\vec{p}$, from the
	parameter space, $\mathcal{P}$, to create an ensemble of simulation
	instances to execute.
In particular,  a {\em sampling manager} takes as input the $k$ aligned input data vectors and samples the possible simulation
			space, balancing computation budget against potential scenario
			fan-out, to decide which $n$ simulation instances to include in its output ensemble: 

	\begin{definition}[Model Simulation Instance]
		A model simulation instance, 
  $\mathbb{E}(\vec{p},\vec{d})$, is a tuple $\langle F, \vec{p},\vec{d}, \mathcal{O}\rangle$, where simulation parameters and input data have been instantiated to $\vec{p}$ and $\vec{d}$ respectively; each execution instance also has a corresponding output data instance $\vec{o}$. 
	\end{definition}
	The  $n$ selected  simulation instances (i.e., the local simulation ensemble to be executed) are passed to an {\em execution manager}, which
executes the simulator configurations marked for ensemble inclusion within a
			cluster, allocating and load balancing instances as necessary. 
 Once the simulation instances have been
		executed, the resulting data are collected or aggregated before being passed to the downstream DS-Actor(s).

	\begin{figure}[t]
		\begin{tabular}{p{0.98\linewidth}}
			\centerline{\includegraphics[width=.8\linewidth]{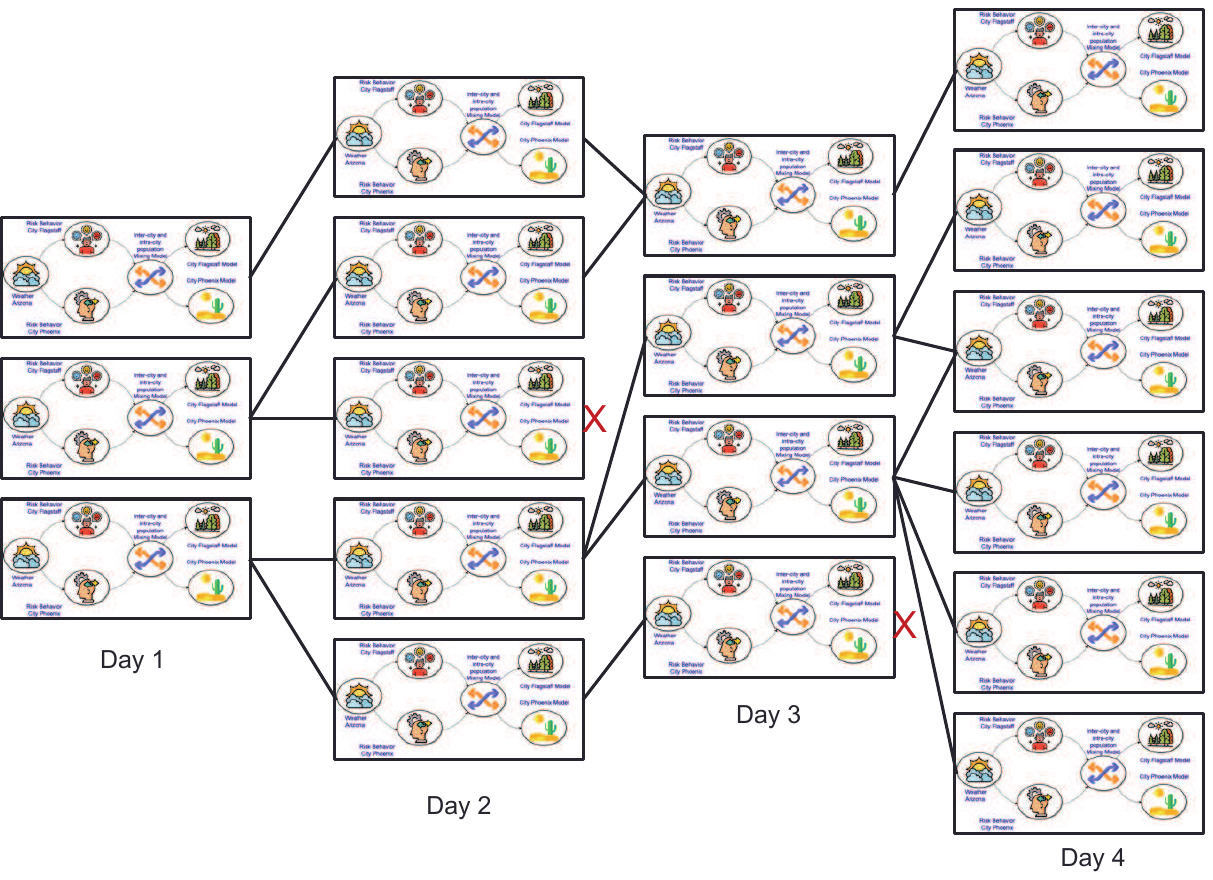}} \\
{(a)			A provenance graph encodes all simulation instances in an ensemble. As simulation ensemble advances, DataStorm adaptively decides which new simulations to execute: some simulations instances may lead to multiple new scenarios, whereas some may be dropped by the sampling manager}\\\\
			\centerline{\includegraphics[width=.8\linewidth]{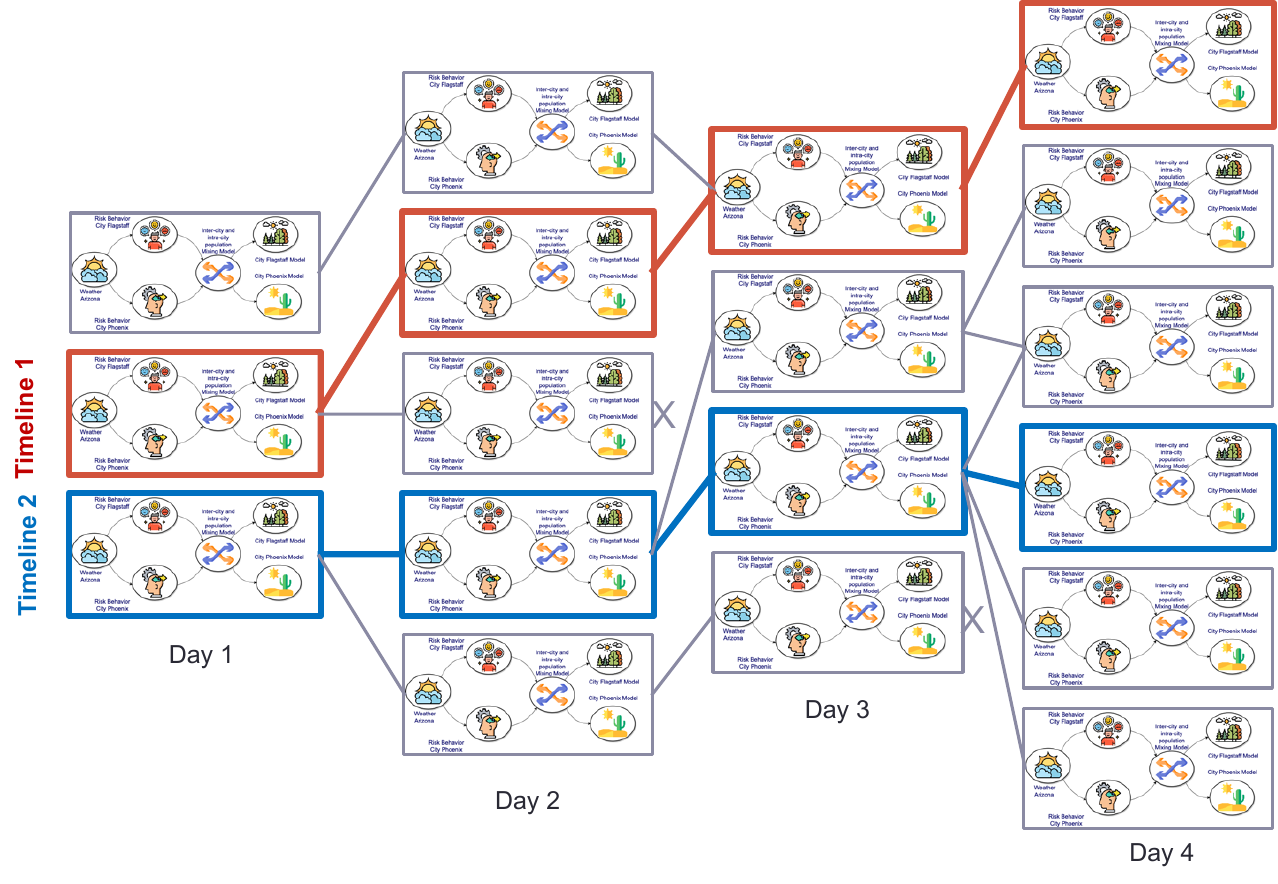}}\\
			{(b) Two timelines extracted from the ensemble: DataStorm-EM helps the user select and explore alternative timelines, each corresponding to a possible scenario}
		\end{tabular} 
		\vspace*{-0.15in}
		\caption{(a) Provenance captures the complete
			histories of simulation instances in the ensemble; (b) a
			timeline on the other hand captures a
			maximal and consistent (i.e. for each given time
			instant, $t$, there is at most one
			simulation instance of each model type) subset
		}\label{fig:provenance}
					 \vspace*{-0.2in}
	\end{figure}

\subsection{Ensemble Graph, Provenance, and Timelines}

Through the continuous, coupled execution of a workflow, DataStorm creates a simulation ensemble which captures multiple  potentially-plausible scenarios, each representing one possible way that future events play out. 
 The overall ensemble consists of multiple local model simulation instances that are stitched together into an {\em ensemble graph}, based on the input-output relationships, reflecting both temporal and model dependencies of the 
 instances.

\begin{definition}[Ensemble Graph]        
An ensemble graph $G_e(V_e,E_e)$ is a graph, where the nodes correspond to the individual model simulation instances and edges represent the data exchanged between them.
Each node $v_i \in V_e$ is labeled with a $\langle m_i, p_i, d_i, w_i, o_i \rangle$, indicating the model	$m_i$ that has been executed using input parameters $p_i$ and data $d_i$, and generating output $o_i$ corresponding to a timewindow $w_i = [t_{start}, t_{end}]$. Each edge $e_j \in E_e$ is labeled with the data being exchanged between the two end nodes.
\end{definition}

\noindent Intuitively, the ensemble graph captures the complete
	history (or {\em provenance}) of each executed model simulation instance and 
	collectively encodes the parameters, assumptions, and
	predictions that led to the execution of that particular
	instance (Figure~\ref{fig:provenance}(a)) and, consequently, weaves and encodes, in a compact form, alternative scenarios (or {\em timelines}) embedded in the entire simulation ensemble. 

Note that, as we see in Figure~\ref{fig:provenance}(a), for a given time instant, $t$, we can have multiple instances of a given model  providing alternative outcomes for that instant. Therefore,  before they can be displayed, explored, and analyzed, the {\em relevant} individual timelines encoded in the ensemble graph need to extracted. 
	We  define a {\em timeline} as a \underline{maximal}
	subset of an {\em ensemble graph} ($G_e$), such that for each
	given time instant, $t$, and for each model $m$, there is at
	most one simulation instance of that model type whose window
	$w$ covers the time instant $t$
	(Figure~\ref{fig:provenance}(b)). Intuitively, each {\em
		timeline} corresponds to a different causal history and an
	{\em execution provenance} may contain many such histories,
	each of which can be separately displayed. The set is {\em
		maximal} in the sense that (a) the simulation time instants
	that are not covered by simulation instances of all models are
	minimized and (b) the causal relationships (encoded as
	input-output pairing among simulation instances) are maximally
	preserved (i.e., if a simulation instance is in the timeline,
	other instances that  produced data that are
	used as input by this instance are also 
	in the timeline).
	
	Since the number of timelines encoded in an ensemble graph can grow exponentially with the number of parameters sampled and included in the local simulation ensemble at each execution step, {\em DataStorm-EM} provides (a) algorithms to help explore the timelines embedded in a simulation ensemble through efficient extraction algorithms that not only maximize user supplied preference criteria, but also help diversify the extracted timelines, (b) ensemble visualization tools to explore these timelines, the underlying model parameters, and the outcomes, and (c)  tools that help analyze the resulting ensembles of multi-variate time series~\cite{xm2a, selego} and help discover the underlying causal patterns~\cite{causal}.

\section{Use Scenario}

Dynamics of complex interconnected systems under pandemic and other disaster/emergency scenarios are extremely difficult 
with the  data  technologies~\cite{10.1145/3206025.3206047,6012029} available today, yet decision makers need ways to estimate the implications of actions they are considering.
Several simulation software have been developed within epidemics, such as GLEaM~\cite{van2011gleamviz}, STEM~\cite{STEM}, and PanVax~\cite{panvax}, as well as biosurveillance and threat assessment tools, such as IBIS and BSVE. Yet, a complex disaster process, such as the COVID-19 pandemic, cannot be simulated through a single monolithic model and an integrated data and model framework that supports complex, coupled system
simulation ensembles
to improve our ability to forecast infectious disease spread, and the effects of interventions, across heterogeneous communities is lacking.
Building on DataStorm (and specifically {\em DataStorm-EM}), we are developing a PanCommunity platform which aims to support seamless integration and manipulation of independently developed, reusable component models and analysis components for improving pandemic response.
Figure~\ref{fig:modeldiagram} shows a sample multi-model integrated pandemics scenario, integrating models necessary to analyze the impacts of various environmental and behavioral factors and their inter-dependencies.
The six pandemic-related models covering Arizona weather, Phoenix behavior, Flagstaff behavior, a mixing model, Phoenix city conditions and Flagstaff city conditions:  The weather model predicts the weather of all the regions at a State level. The behavior models predict the behavior of the population of a certain city  based on the spatio-temporal outputs of the weather model as well as the current state of the disease at that city. The mixing model combines behavioral data into inter-city and intra-city mixing rates. Finally, the city models use the mixing matrix to run city-level epidemic simulations using delay differential equations~\cite{azad2022sirtem}. 
Each of the components described above and visualized in Figure~\ref{fig:modeldiagram} are implemented as a DataStorm actor.

\begin{figure}[t]
    \centering
    {\includegraphics[width=.42\textwidth]{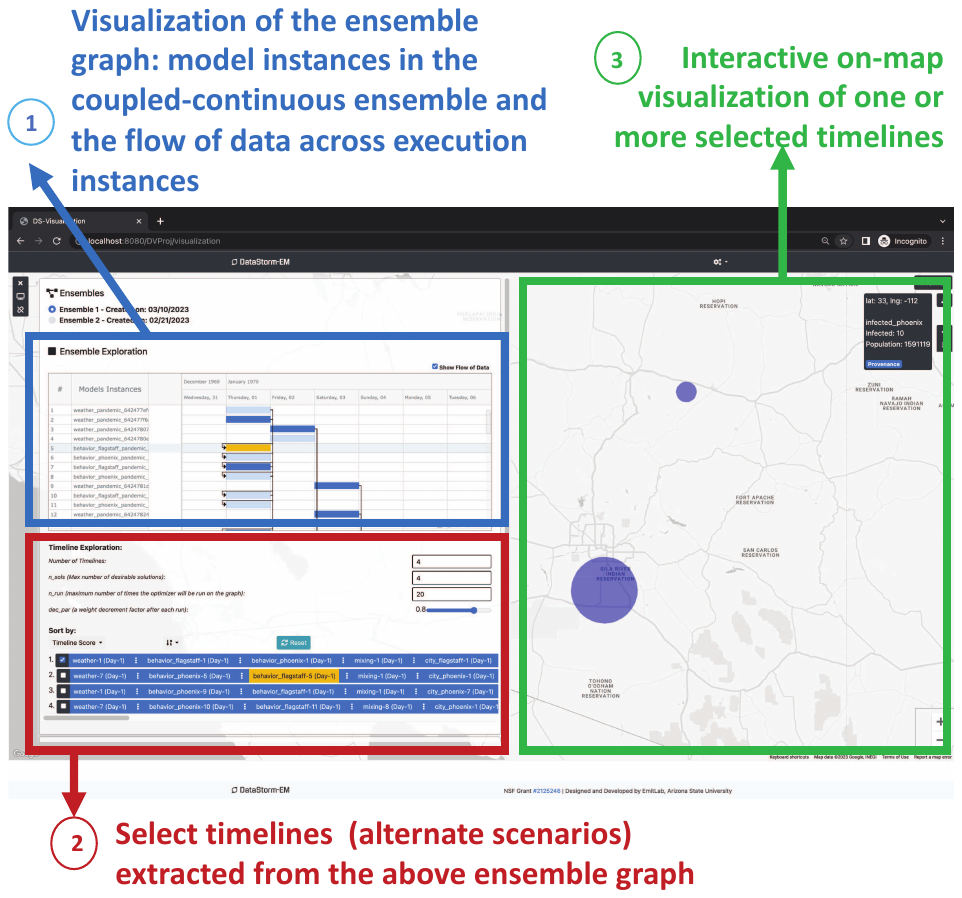}}
    \vspace*{-0.1in}
    \caption{{\em DataStorm-EM} timeline exploration  interface}
    \label{fig:demo}
    \vspace*{-0.15in}
\end{figure}

DataStorm uses an optimization framework to select most productive samples across  weather, behavior, and disease parameters to generate an ensemble of potential timelines, each represented in the form of a multi-variate time series depicting how the susceptible, exposed, infected and recovered populations change over time. The various models sample different parameters throughout the simulation. The weather model samples the temperature where as the behavior models sample the transmission rate, the contact rate, the risk averse or the risk tolerant parameters.
The resulting ensemble, weaving together alternate scenarios, is then represented in an ensemble graph as described in the earlier sections. 
During the demonstration, {\em users will be able to explore alternative (consistent, yet diverse) timelines weaved within ensemble simulations} and make causal inferences regarding the interplay between various weather, behavior, and disease parameters (Figure~\ref{fig:demo}). 

\vspace*{-0.1in}
\section{Conclusions}
In this paper, we have introduced the novel {\em DataStorm-EM} platform for simulation ensemble management, optimization, analysis, and exploration. DataStorm-EM, complements our prior work on continuous-coupled simulation ensemble generation (DataStorm-FE) and   enables decision makers to effectively and efficiently plan their next course of action through large data- and model-driven simulation ensembles. DataStorm-EM enables the decision makers to save data related to a specific timeline extracted from the continuous-coupled simulation ensemble for further analysis and interpretation.
 
\section*{Acknowledgements}
We thank Profs. Guilia Pedrielli and Gerardo Chowell for their contributions to the PanCommunity project.

\bibliographystyle{ACM-Reference-Format}
\bibliography{main}

\end{document}